\documentstyle[12pt]{article}

\begin{document}
\newcommand{\an}{\alpha_s^{naive}}
\newcommand{\aex}{\alpha_s^{exact}}

\begin{center}

{\Large \bf Renormalization schemes and renormalons}

N.V. Krasnikov and A.A. Pivovarov

{\it Institute for Nuclear Research of the Russian Academy of Sciences,
Moscow 117312}

\end{center}
\centerline{\bf Abstract  }

\noindent
We describe some ways how higher order corrections can reveal
themselves if integrated over the infrared region. We show that in
different 
renormalization group (RG)
schemes and for some observables one has no factorial
divergences.
We argue that for treating things in the infrared region it is
preferable to start with a RG scheme without the infrared Landau
pole in the running coupling constant.
The uncertainties for the $\tau$
lepton width resulting from accounting for higher
order corrections are discussed.

\vskip 1cm

With new
high order corrections of perturbation theory hardly             
available anymore 
in cases like $e^+e^-$ annihilation or $\tau$ lepton
width \cite{chet}
it is tempting to speculate on the general
structure of series within 
perturbation theory (PT) \cite{thooft,david,almul}.
Some attention has been recently paid to possible factorial
divergences in PT series for observables that include integration
over an infrared region in momentum space \cite{zakh,beneke,bal,neub}.
At the level of diagrams within 
PT it happens to any observable
and the factorial divergence due to the simple bubble chain 
diagrams is
predicted. In fact this prediction is not well justified 
in QCD because of
the choice of a particular subset of diagrams and even of the
special contributions (nonabelianization \cite{be0}).
Next terms of PT expansion can cancel these divergences because
such terms become large at small momenta
and can not be treated as corrections.
In QED there is a formal parameter for ordering diagrams
within the $1/N_f$ expansion and the statement about the factorial
growth of coefficients can be confirmed by direct computation 
\cite{qed} but
it is known also that QED 
within the $1/N_f$ expansion
has practically no sense
at all and using it as a guide for general structure 
of QCD is not
well grounded.

In this paper
we consider some observables that are represented by integrals over
the infrared region and give several ways to define them using the
freedom of choice of the RG scheme. The main conclusion we draw is
that the results of integration 
can easily be made well defined without any
nonperturbative (in a strictly defined sense) contributions.
These results are ambiguous 
to the same degree as any ordinary PT series,
numerically it can be important because in the infrared region the
coupling constant becomes large in most of ``natural'' RG schemes.
However it can be made small as well 
by some particular choice of extrapolation to low momenta.

The paper is organized as follows.
First we define our main object as an integral of 
some PT expansion over infrared region and show that
the corresponding expression can contain 
no factorials depending on
the integrand. Even with the simplest version of RG analysis
including only the leading $\beta$ function and summation of leading
log's, the Landau ghost contributions to different orders in
$\alpha_s$ can cancel each other. This leads us to conclusion
that infrared renormalons or factorial growth of coefficients
due to
some infrared integration has no invariant meaning.

Next we present a number of examples where integrals are well
defined and discuss their properties as functions of the
boundary value of the coupling 
$\alpha_s(s)$. It is shown that integrals can be
analytic or nonanalytic in $\alpha_s(s)$ 
in the vicinity of $\alpha_s(s)=0$ depending on a
$\beta$ function used. So, extrapolation of PT evolution to low
energy domain can be done in different ways and the results vary
strongly.

These two parts accomplish our theoretical consideration of the
problem with conclusion that nothing definite can be said about
the structure of higher order terms on the basis of some subset
of diagrams. 

Then we go to practical applications and estimate uncertainties
of theoretical predictions for 
the $\tau$ lepton width when different approaches for defining the
integration over infrared region are used.
This problem has been widely discussed in the literature 
(as recent references see, e.g. \cite{alt,kat,neub1})
so we limit ourselves to qualitatively different versions
of changing RG schemes only. 
The basic (or naive) scheme is one
where only self-consistent number of terms of 
the expansion in the
coupling constant is kept. Note that
adding all terms proportional to 
RG log's can produce any answer -- just the
reflection of the Landau singularity. Then we use the K scheme 
\cite{gru,krasn,katkrpiv}
where there are no corrections to $R(s)$ and the corresponding
$\beta$ function 
in three-loop approximation
acquires an infrared fixed point so that the
coupling constant can be extrapolated to the origin
without any singularity.

Finally we propose a set of schemes that regularize the infrared
behavior of the coupling constant 
in general and allow one to use any reference scheme for
high energy domain.
All these schemes are legal and perturbatively equivalent at high
energies. The uncertainties that come from low energy region
are quite essential as our study shows.

Thus the main object of our interest is the expression
of the form 
\begin{equation}
F(s)={1\over s} \int_0^s\rho(t)\omega(t)dt
\label{mainobject} 
\end{equation}
where $\rho(t)$ is given by a PT expansion
$$
\rho(t)=\alpha_s(t)+\ldots
$$
while $\omega(t)$ is a kinematical factor smooth in the integration
region. Expression (\ref{mainobject}) enters the finite energy 
sum rules analysis \cite{fesr}.
We do not dwell upon the peculiarities connected with the analytic
continuation 
\cite{anal,tau}
and consider eq.~(\ref{mainobject}) as a formal expression.
Note that the formulation with the help of integration
along the circle in the complex
plane does not help much by itself 
because RG improved expressions for
PT approximations of physical quantities have wrong analytical
properties 
so the spurious singularities give contributions to physical results
\cite{ren1}.

We restrict ourselves to the case of $\omega(t)=1$ and compare
$$
F(s)={1\over s}\int_0^s\rho(t)dt
\qquad{\rm and }
\qquad \rho(s)=\alpha_s(s)+\ldots
$$
For technical convenience 
we introduce a new notation for the coupling constant
$
a(s)=\beta_0\alpha_s(s)
$ 
so that
asymptotically 
$a(s)\rightarrow(ln(s/\Lambda^2))^{-1}$
when $s\rightarrow\infty$. 
A RG equation for our rescaled coupling constant $a(s)$ reads
\begin{equation}
s{d\over ds}a(s)=\beta(a)=-a^2(1+ca^2+c_1a^3+\ldots)
\label{RGequ}
\end{equation}
with rescaled values of coefficients $c=64/81$, $c_1=3863/4374$
for $n_f=3$ in the $\overline{\rm MS}$ renormalization scheme
\cite{befunc}.

Taking $\rho(t)=a(t)$ we find 
the factorial growth of coefficients of $a(s)$ expansion of $F(s)$
that reflects the presence of infrared Landau ghost 
(or divergence of the formal
series for the integrand)
\begin{equation}
 F(s)={1\over s}\int_0^s a(t)dt=a(s)\sum_{n=0}^\infty a^n(s)n!,
\quad a(s)>0
\label{formalint}
\end{equation}
where in the leading approximation 
\begin{equation}
a(t)={a(s)\over 1-a(s)ln(s/t)}=a(s)\sum_{n=0}^\infty a^n(s)ln^n(s/t).
\label{onelooprun}
\end{equation}
Last equality in eq.~(\ref{onelooprun}) is formal and valid only
within the convergence circle of the power series
$|a(s)ln(s/t)|<1$, the use of this equality
outside the convergence circle leads to all problems of
factorial growth.
Factorials can be canceled by higher order terms in 
the integrand $\rho(t)$.
Indeed, for $\rho(t)=a(t)+ka^2(t)$ we find for a simplified
$\beta$
function 
$\beta(a)=-a^2$
with $c=0$ that corresponds to $n_f\sim 8$
\begin{equation}
{1\over s}\int_0^s \left( a(t)+k a^2(t)\right)dt
=a(s)+a(s)(1+k)\sum_{n=1}^\infty a^n(s)n!.
\label{nofac}
\end{equation}
If $k=-1$ there are no factorials in the expression 
of the form (\ref{formalint}) and only first term survives.
Thus higher order corrections to the integrand
can drastically change the situation with
factorials. It is 
natural because factorials come from the region where
higher order corrections are not small and comparable (even much
larger) than leading contributions.

In general, we find for integrals of the invariant charge given by 
eq.~(\ref{onelooprun})
$$
{1\over s}\int_0^s a(t)^pdt
={a(s)^p\over (p-1)!}\sum_{n=0}^\infty (n+p-1)! a^n(s)
$$
and for $\rho(t)=a(t)\sum_{p=0}^N \rho_p a^p(t)$
$$
{1\over s}\int_0^s \rho(t)dt
=a(s)\sum_{p=0}^N{\rho_p\over p!}\sum_{n=p}^\infty a^n(s) n!
$$
$$
=a(s)\sum_{n=N}^\infty a^n(s)n!\sum_{p=0}^N{\rho_p\over p!}
+a(s)\sum_{p=0}^{N}{\rho_p\over p!}
\sum_{n=p}^{N-1} a^n(s) n!.
$$
The last addendum is a finite sum 
(an analog of $a(s)$ in eq.~(\ref{nofac}))
while the coefficient in front of 
the infinite sum with factorial terms is a linear combination 
made from the coefficients $\rho_p$ of the observable $\rho(t)$.
Inclusion of
nonvanishing coefficients of higher order terms of the $\beta$ function
is straightforward and only leads to extended combinatorics leaving the
statement about the factorial growth unchanged 
if a $\beta$ function keeps to be negatively defined 
for all $a$.
Even if the factorial growth persists (not complete cancelation
between contributions from different orders) its
coefficient (the strength)
is determined by a linear combination of coefficients $\rho_n$
coming from all orders. It is a consequence of the fact that at low
energy the formal ordering with $a(s)$ is invalid and all orders
contribute a comparable amount to the coefficient of factorial.
In other words, factorials appear due to RG log's but in higher orders
coefficients of these log's contain not only the $\beta$ function
coefficients but lower order terms of observables themselves.
So the behavior at large orders of PT for the integral 
can not be fixed without knowing the
whole series for the $\rho(t)$ that is hopeless.
The finite series for a $\beta$ function in schemes like 
't Hooft's one 
is illusion -- it is still a
nonperturbative statement about vanishing coefficients in all orders.

Now we go over to examples without the Landau ghost in the infrared
region that are free of any problems with integration or factorial
growth.
The purpose can be achieved with different means.
In higher orders a $\beta$ function can have a fixed point
in some scheme. This happens for $e^+e^-$ annihilation in third order
in K scheme without corrections.
Or one can write down some new $\beta$ functions that 
are
defined at all values of coupling and determine its evolution
for all momenta in a smooth way.

We start with the second option and consider the model $\beta$
function \cite{ren1}
\begin{equation}
\beta(a)=-{a^2\over 1+\kappa a^2}, ~~\kappa>0
\label{beta}
\end{equation}
while $\beta^{as}(a)=-a^2+\ldots$
Then the RG
equation
(\ref{RGequ})
has a solution
\begin{equation}
a(z)
={-ln{z\over \Lambda^2}
+\sqrt{ln^2{z\over \Lambda^2}+4\kappa}\over 2\kappa}
\label{simplesolution}
\end{equation}
and the unphysical pole at $z=\Lambda^2$ of the
asymptotic solution $a^{as}(z)=(ln(z/ \Lambda^2))^{-1} $
disappears.
Thus, the particular way of summing an infinite number of
specific perturbative terms for the $\beta$
function can cure the Landau pole
problem. Let us stress that there are no nonperturbative terms added
but the freedom of choosing a renormalization scheme for an infinite
series was used instead. This is however beyond the formal framework
of perturbation theory where only finite order polynomials in the
coupling constant are allowed as expressions for any quantity.

Formula (\ref{beta}) can be considered
either as a pure PT result in some particular
RG after an infinite resummation
or as a sort
of Pade approximation of some real $\beta$ function that might
include
nonperturbative terms as well.
The only important point for us here is that the running coupling
obeying the RG equation with such a
$\beta$ function has a smooth continuation to the infrared region.

The effective charge given by eq.~(\ref{simplesolution}) 
has a correct asymptotic
behavior at $z \rightarrow \infty$ and no singularity in the whole
complex plane with a cut along the positive semiaxis if one considers
it as a running coupling constant in Euclidean space.
These properties make it a good expansion parameter for physical
observables that obey the dispersion relation because it has
a proper analytic behavior unlike the asymptotic charge with the Landau
pole. Note that this scheme does not solve the problem
of strong coupling and even fails to bypass this
problem when the expansion for an observable is
alternate and higher order terms can make it negative
at small enough $s$ that can contradict spectrality 
for observables of the type of cross-sections. 
Because the expansion parameter becomes large in infrared region 
the polynomial approximation is invalid in this domain.
Here we encounter a particular case of the
general situation that the expansion in unphysical parameter
$\alpha_s$ is incorrect and the proper way of action is to expand one
physical quantity through another. Such formulation of PT is
physically more justified. The use of K scheme is a particular example
of this approach.

Now we turn to the consideration of 
the structure of expressions of the form
(\ref{mainobject}) with different extrapolation of the running
coupling to infrared region. In general, 
the RG equation 
for 
$$
F(s)={1\over s}\int_0^s a(t)dt
$$
has the form
\begin{equation}
F(a)+\beta(a)F'(a)=a
\label{example}
\end{equation}
and the solution of this differential equation
can be found with quadratures because it is a linear differential
equation of the first order.
At simple $\beta$ functions like $\beta(a)=-a^2$ the
solution of the RG equation (an integral curve) 
fixed by the boundary
condition $a(s)\rightarrow 0$ at $s \rightarrow \infty$
can not be continued to the point $s=0$. 
The integral curve goes only until $s=\Lambda^2$
where there is a vertical asymptote. 
New information is necessary for continuation of the
solution to the origin.
Consider now an example with
\begin{equation}
\beta(a) = {-a^2\over 1+2a},~~~ a>0.
\label{beta2}
\end{equation}
In this case the interval $(0,\infty)$ in $s$ is
mapped uniquely to the interval $(0,\infty)$ in $a$ 
and the solution can be uniquely continued to the origin.
Still a special consideration of small $s$ region is necessary to
determine the correct contribution of the solution of the homogeneous
equation. We find
$$
F(a)=a+a^2+\Delta F(a)
$$
where $\Delta F(a)$
is a solution of the homogeneous equation
\begin{equation}
F(a)+\beta(a)F'(a)=0
\label{homo}
\end{equation}
$$
\Delta F(a)=F_0 a^2 exp(-{1\over a}).
$$
Coefficient $F_0$ is determined from the infrared region to be
$F_0=-1$.
The result is
$$
F=a+a^2-a^2exp(-{1\over a}).
$$
This result can be obtained by explicit integration as well.
Namely, 
the RG equation for the effective charge 
of eq.~(\ref{beta2}) is given by
$$
ln(s/\Lambda^2)={1\over a}-2ln a
$$
and $F(s)$ can be rewritten as ($a_0\equiv a(s)$)
$$
F(s)\equiv F(a_0)=e^{-({1\over a_0}-2ln a_0)}\int^\infty_{a_0}
ae^{({1\over a}-2ln a)}({1\over a^2}+2{1\over a})da
$$
$$
=e^{-({1\over a_0}-2ln a_0)}\int_0^{1/a_0}
e^{\xi}(\xi+2)d\xi
=a_0+a_0^2(1-e^{-{1\over a_0}}).
$$
The last term gives the ``condensate'' contribution to
$F(s)$ at $a_0\rightarrow 0$ $(s\rightarrow \infty)$. Up to
logarithmic corrections 
$$
F^{cond}(s)\sim {\Lambda^2\over s}.
$$
At small 
$s$ the effective charge runs as
$
a(s)\sim \Lambda/\sqrt{s}$
and 
$$
F(s)\sim -{2\Lambda\over \sqrt{s}}.
$$
Note that the $\beta$ function (\ref{beta2}) does not 
regularize the whole physical quantity: higher order terms in
the expansion (like $a^2$) can be nonintegrable. In this
particular example it is the case.
 
This simple example shows that for 
a reasonable definition of the evolution
in the infrared region factorials are absent in the expansion.
The last term cannot be detected if the equation (\ref{example})
is integrated by
series near $a=0$. Indeed,
for $F(a)=\sum_{k=1} f_k a^k$ one has for eq.~(\ref{example}) with $\beta$
function (\ref{beta2})
$$
f_1=1,\quad f_2=1,\quad f_k-(k-3)f_{k-1}=0,\quad k>2.
$$
Thus, $f_k=0$, $k>2$. Note that the last equation would lead to
factorial growth but instead all higher order coefficients vanish.

To study higher orders of PT the Borel transformation is often used.  
The Borel analysis for this example goes as follows
\begin{equation}
F(a)=\int_0^\infty e^{-{\xi\over a}}B(\xi)d\xi
\label{bordef}
\end{equation}
and the Borel image $B(x)$ is 
\begin{equation}
B(x)=1+\theta(x-1)+x\theta(1-x).
\label{borim}
\end{equation}
So the PT series with all coefficients known still
does not
allow to restore the exact answer through Borel
summation.
A naive Borel image for polynomials from definition 
(\ref{bordef}) behaves as $B(x)= \sum_{k=1}f_k x^k/(k-1)!$ that is correct
at small $x$ and completely wrong at large $x$ and has
nothing to do with the exact one (\ref{borim}).
Being rewritten as 
\begin{equation}
B(x)=1+x + (1-x)\theta(x-1)
\label{borim1}
\end{equation}
the result shows that Borel image is singular 
(nondifferentiable) at $x=1$
though this is not a pole singularity.

The corresponding example with $\beta$ function (\ref{beta})
can be written as follows
$$
F(s)=a+{\kappa\over 3}a^3
={1\over s}\int_0^s(a-a^2+{\kappa\over 3}a^3)ds.
$$
The RG equation of the type (\ref{example}) 
can be integrated by series but the homogeneous
equation (\ref{homo}) has no solution satisfying
the boundary conditions at small $s$
(constraint on the growth). 
To understand this feature we note that at small $s$ the      
running coupling behaves like 
$a(s)\sim \kappa^{-1}ln(\Lambda/s)$ and after integration the
logarithmic behavior survives. As for the solution of the
homogeneous equation it contains a factor 
$$exp(-{1\over a}+\kappa a)$$ that is not logarithmic at small
$s$ ($a\rightarrow \infty$) and can not appear in the
resulting expression.
In contrast to the previous example all powers of 
the effective coupling are
integrable here. The rate of increase is however quite large for
higher order terms.
   
Now we define a set of schemes that provide a smooth
extrapolation to the infrared region that are a direct generalization
of the model (\ref{beta2}). The $\kappa$ scheme is
determined by the $\beta$ function
\begin{equation}
\beta_\kappa(a)={ \beta(a)\over 1-\kappa a^n \beta(a)}
\label{kappa}
\end{equation}
where $\beta(a)$ is a $\beta$ function in a reference scheme,
$\overline{\rm MS}$ 
for instance. 
The form (\ref{kappa})
is chosen because of practical
convenience only -- it requires no more work as corresponding
reference scheme and eliminates the Landau pole in infrared region.

The $\beta$ function given by eq.~(\ref{kappa}) is bounded at large
$a$ and eq.~(\ref{RGequ}) has a solution for $a(z)$ that is
defined on the whole positive semiaxis and is free of the Landau
pole.
The absence of singularities (Landau ghost) 
allows one to use the evolution 
of coupling till the origin and destroys the
mathematical part of the reasoning of ref. \cite{zakh} about Borel
nonsummability and also Parisi's conjecture \cite{Parisi} looks wrong in
this particular scheme.

The solution of the type 
(\ref{simplesolution}) 
can be rewritten without using $\Lambda$ in a more
traditional
form through some intermediate energy scale $\mu$
\cite{ren1}. The
parameter $\Lambda$ has no special meaning anymore as a position of
the pole of the running coupling.
Thus, in the pure PT framework one can eliminate
the Landau pole of the coupling using the freedom of the
renormalization scheme choice. The smooth coupling however does not
require any nonperturbative effects to regularize the integrals in
which it appears.

The solution to the RG equation for the invariant charge in the
$\kappa$
scheme is simple because it is closely related
to 
the $\overline{\rm MS}$
running coupling
\begin{equation}
ln(s/\Lambda^2)=\Phi(a)-\kappa {a^{n+1}\over n+1}
\label{phidef}
\end{equation}
where
$$
\Phi(a) \sim \int^a{dx\over \beta(x)}
$$
and the normalization is chosen in the form
\begin{equation}
\Phi(a)={1\over a}-c\ln({1\over a}+c)
+\int_0^a\left({1\over \beta(\xi)}
-{1\over \beta_{(2)}(\xi)}\right)d\xi
\label{phinorm}
\end{equation}
with $\beta_{(2)}(a)=-a^2(1+ca)$.
This gives the standard definition of the parameter $\Lambda$
to be $\Lambda_{\overline{\rm MS}}$.
Two charges are connected through
$$
a_{\kappa}=a-{\kappa\over n+1}a^{n+3} + o(a^{n+3}).
$$ 
The exact connection can be found from eq.~(\ref{phidef}).
So, for $n>1$ they coincide for all observables because in
practice there are no calculation beyond the third order.

Still if the corresponding expansion 
of a physical observable is alternative, higher
order corrections can make the physical quantity negative in some
part of the integration region. In this sense the approach does not
give the full regularization. 

Concluding the theoretical part of the paper we state that there are
different ways to bypass the problem of factorial growth of
coefficients for integrals from physical observables. The simple change
of RG scheme within PT in high energy region can allow a smooth
extrapolation of the effective charge into low energy domain that
makes all integrals well defined. The change of scheme leads to
numerical uncertainties in prediction that is a common feature of any
PT result. We now consider these uncertainties for 
predictions of the $\tau$ lepton
width or for the parameter $\alpha_s(m_\tau^2)$ extracted from
experimental data on this width.
We stress that we limit ourselves only to ``qualitatively'' different
schemes that describe different infrared behavior of an invariant
charge.
We do not discuss small variation of schemes within the qualitatively
equivalent class.

The expression for the $\tau$ lepton width
has the form \cite{bra} 
\begin{equation}
R_\tau=\int_0^{m_\tau^2}{ds\over m_\tau^2}2(1-{s\over
m_\tau^2})^2(1+2{s\over m_\tau^2})R(s)
\label{tautheor}
\end{equation}
with $R(s)$ given by 
$$
R(s)=3(1+{\alpha_s(s)\over \pi}+\ldots).
$$
Using the experimental value (as a recent reference, see \cite{alt})
\begin{equation}
R_\tau^{exp}=3.645\pm0.024
\label{tauexp}
\end{equation}
and writing
$$
R_\tau=3(1+{\pi\over \beta_0}r_\tau)
$$
we find
$$
r_\tau=a(\mu^2)+\dots
\quad {\rm and}\quad r_\tau^{exp}=.4838\pm0.018.
$$
First, 
for the sake of completeness and 
for purposes of normalization we compute the
naive RG result that corresponds to keeping only three terms of
expansion in $a(\mu^2)$ with explicit RG logarithms.
Having rewritten $R(s)$ as
$$
R(s)=3(1+{\pi\over \beta_0}r(s)),
$$
one gets for properly normalized QCD contribution
(we do not discuss any corrections due to electroweak interactions and
take this process only for demonstration of theoretical uncertainties
stemming from the choice of RG scheme in QCD)  
\begin{equation} 
r(s)
=a(\mu^2)+a^2(\mu^2)(k_1+L)
+a^3(\mu^2)(k_2+L(c+2k_1)+L^2)
\label{rnaive}
\end{equation} 
with
$k_1=0.7288$, $k_2=-2.0314$ \cite{gorishn,levan},
$L=\ln{\mu^2\over s}$.
Introducing normalized moments
$$
r_N=(N+1)\int_0^{m_\tau^2}{ds\over m_\tau^2}
\left({s\over m_\tau^2}\right)^N\rho(a)
$$
we get the well known result 
$$
r_\tau=2(r_0-r_2+r_3/2)
$$
and
$$
(N+1)\int_0^1 x^N (-ln x)^p dx ={p!\over (N+1)^p}.
$$
This gives
\begin{equation}
r_N
=a+a^2\left(k_1+{1\over N+1}\right)
+a^3\left(k_2+{c+2k_1\over N+1}+{2\over (N+1)^2}\right).
\label{mom}
\end{equation}
Collecting all together we get the result
\begin{equation}
r_\tau
=a+(k_1+\delta_1)a^2+(k_2+(c+2k_1)\delta_1+\delta_2)a^3
\equiv a+k_1' a^2+k_2' a^3
\label{resultth}
\end{equation}
where $\delta_1=19/12$, $\delta_2=265/72$.
The equation for determination $\alpha_s(m_\tau^2)$
reads
\begin{equation}
r_\tau=r_\tau^{exp}=0.4838\pm0.018.
\label{equat}
\end{equation}
We have two options here: to solve equation (\ref{equat})
with the theoretical expression for $r_\tau$ given by
eq.~(ref{resultth}) 
exactly or perturbatively.
Exact answer is    
$$
a=0.253\pm 0.006 \quad {\rm and}
\qquad \alpha_s(m_\tau^2)=0.353\pm 0.008.
$$
Perturbative method gives
$$
a(m_\tau^2)=r_\tau-k_1' r_\tau^2 - (k_2'-2(k_1')^2)r_\tau^3+O(r_\tau^4)
$$
$$
=r_\tau-(k_1+\delta_1)r_\tau^2
-(k_2+(c+2k_1)\delta_1+\delta_2-2(k_1+\delta_1)^2)r_\tau^3
+O(r_\tau^4)=0.564\pm0.046
$$
and 
$$
\alpha_s(m_\tau^2)={4\pi\over 9}a(m_\tau^2)=0.787\pm 0.063.
$$
Both results are purely perturbative 
and the errors reflect only the experimental
error with the theoretical procedure being strictly fixed.
We do not consider any options with choice of RG without introducing
any qualitative change. We also do not discuss any keeping additional
terms known due to RG (like $\alpha_s^n ln^n(\mu^2/s)$).

Our last remark here is to formulate the result in 
a RG invariant way in the spirit of ref.~\cite{dhar}.
The simplest (not unique) way is to express the
integral through the value of integrand taken at the
end point. If $r(s)=a+k_1 a^2+ k_2 a^3$
then
\begin{equation} 
{1\over s}\int_0^s r(t)dt = r(s)+r(s)^2+(c+2)r(s)^3
\label{RGinv}
\end{equation} 
and for the $\tau$ lepton width we find
\begin{equation} 
r_\tau
=r(s)+\delta_1 r(s)^2
+(\delta_1 c+\delta_2) r(s)^3
=r(s)+{19\over 12}r(s)^2
+\left({19\over 12}c+{265\over 72}\right) r(s)^3.
\label{RGinvtau}
\end{equation} 
Eq.~(\ref{RGinvtau}) corresponds to the spirit of PT
in that respect
that it gives the direct relation between two measurable quantities.
As for numerical values, the exact solution is
$r(m_\tau^2)=0.270$ while the perturbative one is
$r^{PT}(m_\tau^2)=0.123$.
For corresponding couplings we find 
$\alpha_s^{PT}(m_\tau^2)=0.164$ and 
$\alpha_s^{exact}(m_\tau^2)=0.359$. 
So, these two observables are poorly connected by 
PT relation that means that the operation of integration
disturbs the PT expansion for $r(s)$ strongly. 
From formula (\ref{RGinv})
and integration rules for log's we deduce that large
contributions of infrared region should be suppressed,
i.e. only computation of high order moments are allowed
as a PT operation.
For three orders of coupling constant expansion and for 
the $Nth$
order moment we require that the change of coefficients would be
small enough 
\begin{equation}
r_N(r)=r+{1\over N+1}r^2
+\left({c\over N+1}+{2\over (N+1)^2}\right)r^3.
\label{momall}
\end{equation}
This relation (\ref{momall}) does not include any coefficients
$k_i$ of the expansion of $r$ in unphysical parameter
$\alpha_{\overline{\rm MS}}$ and reflects only the influence
of integration. Requiring this influence 
to be perturbative we get $N>1-2$ for particular values of $r$ and
$c$.
The estimate depends on the number of RG log's kept for the expansion:
the more log's kept in the expansion the larger the low limit for 
moments remaining perturbatively related to the integrand. 
This example again shows that the MS
coupling constant is artificial (and auxiliary) quantity. 
And with integration over regions where PT series is invalid
the integration measure can not be chosen arbitrary.
Of course, numerically this particular computation can be made
more appropriate by choosing some intermediate point $s^*$ for
normalization instead of the end point and by using $r(s^*)$,
$s^*<m_\tau^2$ as an expansion parameter for the integral.
This method would correspond to one of ref.~\cite{brod,neub4} but 
for fixed order of moment $s^*$ approaches to $\Lambda^2$ if higher
orders of log's would be kept.   
Still, such a formulation of the calculation is more physical one
because it connects two measurable quantities directly without
intermediate things like $\alpha_{MS}$. The validity of PT expansion
in this case means that two quantities commensurate
and can be a measure for each other.
We see that integration through infrared region violates this property
unless the weight function is properly chosen. The choice depends also
on the order of PT:
at higher orders the suppression must be larger.
So, the result of integration and the integrand are
poorly connected by PT. If we had only higher moments
(say, third one) for the width 
the situation would be different for this particular
order of PT.  

In fact, the most successful method in this framework 
would be 
the Stevenson's approach \cite{pms0,pms}
that requires an independent optimization for 
the integral as an independent PT quantity. 
Being extremely flexible it creates the best scheme for every process
and every order of PT independently. We did not check however whether
this method gives the convergent answer for the problem at hand.

Next we consider methods that allow to integrate after applying
the RG that is beyond the scope of conventional PT.

First, in three-loop approximation in 
the K scheme there is a fixed point for the running coupling and
consequently there is no Landau pole problem.
Namely, in the scheme where there are no corrections to $r(s)$ 
the running coupling reads 
$$
r(s)\equiv a_K=a+k_1a^2+k_2a^3
$$
while the $\beta$ function in this scheme is 
$$
\beta_K(a_K)=-a_K^2(1+ca_K+(c_1-ck_1-k_1^2+k_2)a_K^2+\ldots)
\equiv -a_K^2(1+ca_K+c_1'a_K^2+\ldots)
$$
with $c_1'=-2.2552$.
Fixed point is given by the equation
$$
1+ca_K+c_1'a_K^2=0
$$
with the solution
$a_K=0.8637$, and $(\alpha_s)_K=(4\pi/9)a_K=1.206$.
The relation between coupling in K and $\overline{\rm MS}$ schemes is
$$
a_K=a+k_1a^2+k_2a^3,
\qquad a=a_K-k_1a_K^2-(k_2-2k_1^2)a_K^3.
$$
Then the procedure is as follows.
We compute integrals exactly
an find
\begin{equation}
\alpha_s(m_\tau^2)_K=0.380\pm 0.013.
\label{kres}
\end{equation}
Computation of $\overline{\rm MS}$ coupling constant gives two answers
again
$$
\quad\alpha_s(m_\tau^2)^{exact}=0.361\pm 0.014
$$
$$
a(m_\tau^2)^{pol}
=a_K(m_\tau^2)-k_1 a_K^2(m_\tau^2)-(k_2-2k_1^2)a_K^3(m_\tau^2),
$$
$$
\alpha_s(m_\tau^2)^{pol}={4\pi\over 9}a(m_\tau^2)^{pol}=0.392\pm 0.017.
$$
Note that the difference between them is larger than 
the uncertainty due to the 
experimental error and two intervals hardly
overlap. 

Finally, we give the results in our $\kappa$ scheme analysis
that is more general and universally applicable to integration
in any basic scheme as well.
This is a generalization of the fixed point approach.

The spectral density in $\kappa$ scheme 
(with $n=0$) up to third order is
$$
\rho(s)=a_\kappa+k_1a_\kappa^2+(k_2+\kappa)a_\kappa^3,
$$
$$
a_\kappa=a-\kappa a^3,\quad
a=a_\kappa+\kappa a_\kappa^3.
$$
The expansion of $\beta$ function reads
$$
\beta(a_\kappa)
=-a_\kappa^2(1+c a_\kappa+(c_1-\kappa) a_\kappa^2+\ldots)
$$
so introducing 
$\kappa$ is equivalent at high energies to change of $c_1$.
At low energies however they are different.

Integration for moments can be easily rewritten in term of the charge
itself
$$
r_N=(N+1)\int_0^{m_\tau^2}{ds\over m_\tau^2}
({s\over m_\tau^2})^N\rho(a)
$$
$$
=(N+1)\int_\infty^{a_\kappa}
exp(N+1)(\Phi(\xi)-\kappa\xi-\Phi(a_\kappa)+\kappa a_\kappa)
({1\over \beta(\xi)}-\kappa)\rho(\xi)d\xi.
$$
Introducing the variable $\zeta=1/\xi$ we get the practical version
$$
r_N=(N+1)\int_0^{a_\kappa^{-1}}
exp(N+1)(\Phi(\zeta^{-1})-\kappa\zeta^{-1}-\Phi(a_\kappa)+\kappa a_\kappa)
\left({1\over \zeta^2 \beta(\zeta^{-1})}+\kappa \zeta^2\right)
\rho(\zeta^{-1})d\zeta.
$$

Results depend on $\kappa$. This is the ordinary RG dependence
that is strong enough because the accuracy is different at large
momenta where we keep only the expansion and at small momenta where the
exact formulae have to be used to make integrals finite.
The obtained results are given in Table \ref{t:1}.
Here $a_\kappa$ 
is found from integration, $\an$ is found from the naive
(to third order) relation between the schemes
$$
\an={4\pi\over 9}(a_\kappa+\kappa a_\kappa^3),
$$
while $\aex$ is found from exact formulae for RG scheme relations  
(\ref{phidef},\ref{phinorm}).

For finding the parameter $a_\kappa$ from the integral it is
useful to know the derivative of the integral with respect to
the boundary value $a_\kappa(m_\tau^2)\equiv a_0$
$$
{dr\over da_0}=-{1\over \beta_\kappa(a_0)}2(r_0-3r_2+2r_3).
$$

\begin{table}
\begin{tabular}{|c|c|c|c|} \hline
$\kappa$ &$a_\kappa$ &$\an$       &$\aex$     \\ \hline
1.5      &0.2425      &0.368(16)   &0.380(18)  \\ 
1.6      &0.2209      &0.333(09)   &0.342(10)  \\ 
1.7      &0.2118      &0.318(08)   &0.326(09)  \\ 
1.8      &0.2068      &0.311(07)   &0.319(08)  \\ 
1.9      &0.2037      &0.307(07)   &0.315(08)  \\ 
2.0      &0.2016      &0.304(07)   &0.313(08)  \\ 
2.1      &0.2003      &0.303	   &0.312	\\
2.2      &0.1993      &0.303	   &0.311	\\ 
2.3      &0.1986      &0.303	   &0.312	\\ 
2.4      &0.1982      &0.303	   &0.312	\\ 
2.5      &0.1979      &0.303	   &0.313	\\ 
2.6      &0.1977      &0.304	   &0.315	\\ 
2.7      &0.1975      &0.305	   &0.316	\\ 
2.8      &0.1975      &0.306	   &0.318	\\ 
2.9      &0.1975      &0.307	   &0.319	\\ 
3.0      &0.1975      &0.308	   &0.321	\\ 
3.2      &0.1976      &0.310	   &0.325 	\\ 
3.4      &0.1978      &0.313	   &0.329	\\ 
3.6      &0.1980      &0.315	   &0.334	\\ \hline
\end{tabular}
\caption{$\kappa$ scheme results}
\label{t:1}
\end{table}

We can find the derivative RG equation for coupling $a_\kappa$
describing its dependence on the parameter $\kappa$
\begin{equation}
{da_\kappa\over d\kappa}=a_\kappa\beta_\kappa(a_\kappa).
\label{kapeq}
\end{equation}
This is a particular case of RG equations
$$ 
{da\over dc_n}
=-\beta(a)\int_0^a{x^{n+3}dx\over \beta^2(x)}, \quad n\ge 1,
$$
that describe the dependence of the running coupling
on coefficients of the $\beta$ function.
Note that the dependence on $c$ is fixed by the choice of the 
parameter $\Lambda$ to be $\Lambda_{\overline{\rm MS}}$.
The extraction of $a_\kappa$ is done under the assumption of RG
invariance 
of the expression for the $\tau$ lepton width
so the only reliable data for $a_\kappa$
can be taken from that part of Table 1 where
equation (\ref{kapeq}) is satisfied. This equation can be easily solved
analytically 
(it is a linear equation if $\kappa$ is considered as a
dependent function and $a_\kappa$ as an independent variable), 
we have preferred however to solve it numerically in the vicinity of
the value ($\kappa=2.1, a_\kappa=0.2003$). 
The solution does not match well
the pattern of the dependence of the extracted $a_\kappa$ presented in
Table 1 that
means that higher order terms of perturbative expansion 
for the width are essential.
If the theoretical expression for 
$r_\tau$ was exactly independent of the scheme the
extracted charge would satisfy eq.~(\ref{kapeq}). The
actual solution with the initial condition 
($\kappa=2.1$,$a_\kappa=0.2003$)
which is our best estimate is presented in Table 2.
\begin{table}
\begin{tabular}{|c|c|} \hline
$\kappa$ &$a_\kappa$  \\  \hline
1.5      &.2058 \\
1.6      &.2049 \\
1.7      &.2039 \\
1.8      &.2030 \\
1.9      &.2021 \\
2.0      &.2012 \\
2.1      &.2003 \\
2.2      &.1994 \\
2.3      &.1986 \\
2.4      &.1978 \\
2.5      &.1969 \\
2.6      &.1961 \\
2.7      &.1953	\\ \hline
\end{tabular}
\caption{RG dependence of $a_\kappa$ on $\kappa$}
\end{table}
So the correct pattern is around $1.9<\kappa<2.3$ where
the prediction is $\alpha_s(m_\tau^2)=0.312$.
 
Note that contrary to possible impression 
the prediction in fixed point scheme (or in K scheme)
is also
nonstable. Indeed, it is easy to introduce a set of schemes
parameterized with the fixed point value of the invariant charge --
the extracted $\alpha_s(m_\tau^2)$ will depend on the scheme within
the set.
A $\beta$ function for such a set could have the form
\begin{equation}
\beta_f(a)=\beta(a)(1+\kappa\beta(a))
\label{betafixpnt}
\end{equation}
that introduces a dependence on an external scheme parameter $\kappa$.
All schemes of the type (\ref{betafixpnt})
have a fixed point with different value of the coupling
constant depending on parameter $\kappa$.   
  
To conclude,
we have exploited the renormalization scheme freedom to show that
some infinite subsets of diagrams can be
even convergent not just summable
in some generalized way. Then
with lack of any parameter or criterion for
choosing a particular set of diagrams
the renormalization scheme freedom --
extended to partially summed infinite series --
can change the
conclusion about the pattern of behavior of PT
series in high orders dictated by the leading order RG and by the
presence of the infrared pole in the running coupling constant.
The very existence of renormalons can well be
the attribute of a renormalization scheme with the Landau ghost.

In fact, the dependence of the extracted values 
of the coupling constant for the $\tau$ lepton width on 
schemes is rather large because the energy scale ($m_\tau^2$)
is quite low. So, it might be reasonable to fix the
scheme in an arbitrary, and somehow simplest, way 
and then to parameterize the
low momenta region in the integral sense only without
detailed description of the behavior of the running
coupling from point to point. This can be done in terms of
distribution. Adding the localized distribution
like $\delta$ function and its derivatives can make
the integral well defined \cite{ren1,bub}. 
These localized contributions look
as nonperturbative terms.

As for phenomenological applications, e.g.
\cite{big,benbraun,man,web,akh}, there are models where some
extrapolation of the running charge into the infrared
region
is used for practical calculation (normally under the sign of
integration).
In case of the $\tau$ lepton width the uncertainties are large.
Our results show that with an extrapolation chosen one
can not freely add the contribution of gluon condensate to take into
account nonperturbative effects
and "improve" the computation
 -- it must be coordinated with the
continuation of the running coupling constant into infrared region. 
In fact, the fixed gluon condensate corresponds to a
fixed continuation -- the pattern of principal value looks very
attractive in this respect.
An analogous situation
appears in a toy model considered in ref.~\cite{bound}.

Thus, the integration over an infrared region involves the strong
coupling dynamics and is arbitrary to a large extent.
The final expression of the results through the coupling constant at
the boundary value is misleading -- the series is bad defined
in PT and requires some additional inputs concerning the behavior of
the running coupling constant at small momenta. 
The operation of integration must be restricted, it remains valid only 
if the low $s$ are
suppressed by the weight function. Then the result of integration is
perturbatively connected to the integrand and the procedure can be
formulated in a RG invariant way.   

{\bf Acknowledgments.}

One of us (NVK) is indebted to participants of CERN Renormalon
Workshop for very hot discussions that was one of the motivations for
writing this paper.

%\newpage

\end{document}